\documentclass[pra,aps,amssymb,twocolumn,noshowpacs]{revtex4-1}
\usepackage{color}
\usepackage{graphicx}
\usepackage{hyperref}
\usepackage{amsmath}
\usepackage{latexsym}
\usepackage{amssymb}
\usepackage{mathrsfs}
\usepackage{mathtools}
\usepackage{layout}
\usepackage{verbatim}
\usepackage{multirow}
\usepackage{bm}
\usepackage{amsfonts,epsfig}
\usepackage{xcolor}

\begin{document}
\title{Quantum gates with weak van der Waals interactions of neutral Rydberg atoms}

\date{\today}
\author{Xiao-Feng Shi}
\affiliation{School of Physics and Optoelectronic Engineering, Xidian University, Xi'an 710071, China}
\author{Yan Lu}
\affiliation{School of Physics and Optoelectronic Engineering, Xidian University, Xi'an 710071, China}

\begin{abstract}
Neutral atoms are promising for large-scale quantum computing, but accurate neutral-atom entanglement depends on large Rydberg interactions which strongly limit the interatomic distances. Via a phase accumulation in detuned Rabi cycles enabled by a Rydberg interaction of similar magnitude to the Rydberg Rabi frequency, we study a controlled-phase gate with an arbitrary phase and extend it to the controlled-NOT gate. The gates need only three steps for coupling one Rydberg state, depend on easily accessible van der Waals interaction that naturally arises between distant atoms, and have no rotation error in the weak interaction regime. Importantly, they can work with very weak interactions so that well-separated qubits can be entangled. The gates are sensitive to the irremovable fluctuation of Rydberg interactions, but can still have a fidelity over 98\% with realistic position fluctuation of qubits separated over 20~$\mu$m.

\end{abstract}
\maketitle

\section{introduction}
Neutral atoms can be rapidly entangled when they are excited to high-lying Rydberg states, which renders the possibility to use neutral atoms for large-scale quantum computing~\cite{Saffman2010,Saffman2016,Weiss2017,Adams2019,Wu2021youli,Morgado2021}. There have been several experiments demonstrating entanglement between individual neutral atoms by Rydberg interactions~\cite{Wilk2010,Isenhower2010,Zhang2010,Maller2015,Jau2015,Zeng2017,Levine2018,Picken2018,Levine2019,Graham2019,Jo2019,Madjarov2020} primarily via the blockade mechanism~\cite{PhysRevLett.85.2208} where two nearby Rydberg atoms should possess a strong interaction $V$. The value of $V$ drops quickly when the qubit spacing increases, so it is necessary to place the qubits close enough for the blockade condition to hold. For example, the interatomic distance was in the range $[3.6,~5.7]~\mu$m in recent experiments of high-fidelity neutral-atom entanglement~\cite{Levine2018,Levine2019,Graham2019,Madjarov2020}. On the other hand, the yet to be large-scale quantum processor is supposed to host a large number of qubits in which entanglement operations between distant qubits are required in a general computational task. To tackle this issue, Ref.~\cite{Weimer2012} suggested a method to entangle two qubits separated by a chain of ancillary qubits by adiabatically following the many-body ground state of the qubit chain coupled via the Rydberg blockade mechanism, and Ref.~\cite{Cesa2017} proposed to entangle distant qubits by coupling two targeted qubits with a group of ancillary atoms via Rydberg interactions. These methods depend on strong interactions of nearby Rydberg atoms. For example, Ref.~\cite{Weimer2012} analyzed a model by assuming a large enough $V$ enabled by a small average interatomic spacing $1~\mu$m, so that the entanglement of two logic qubits separated by $20~\mu$m would require a coherent control over around 19 ancillary qubits. 

In this work, we analyze an entangling gate between two well-separated neutral atoms by partially exciting them to Rydberg states. In contrast to the gate protocols dependent on the blockade mechanism, we find that a tiny $V$ can rapidly generate entanglement between ground hyperfine levels via detuned ground-Rydberg Rabi oscillations, and a controlled phase gate with an arbitrary phase $\theta$ can be created accurately. The intrinsic fidelity of the gate is perfect when $V$ is frozen and when the Rydberg-state decay is ignored, but for realistic setups where there will be fluctuations of qubit positions~\cite{Graham2019} and Rydberg-state decay, the gate can still have a fidelity $0.99$ with qubit spacing over $20~\mu$m. Compared to the Rydberg gate by a dynamical phase shift $Vt/\hbar$ from the Rydberg interactions~\cite{PhysRevLett.85.2208}, our gate does not require the Rydberg Rabi frequency $\Omega$ to be much larger than $V/\hbar$ and, thus, a high-fidelity implementation is less technically demanding, where $h(\hbar)$ is the (reduced) Planck constant. Compared to the Rydberg gates in~\cite{Shi2017} that work with similar condition $V/\hbar\sim\Omega$ but need five pulses to couple multiple Rydberg states, the gate in this work needs fewer pulses to couple only one Rydberg state. In contrast to previous Rydberg gates based on F\"{o}rster resonances~\cite{Shi2017,Beterov2016,Petrosyan2017,Beterov2018,Beterov2018arX,Khazali2020,Beterov2020,Stojanovi2021}, our gate depends on van der Waals interactions that naturally appear in well separated atoms without resorting to external fields for tuning a F\"{o}rster resonance. Since entangling operations between well-separated qubits are necessary for large-scale quantum computing, our method can simplify the quantum circuit in quantum computing because otherwise many standard Rydberg gates are needed to entangle two distant qubits.

The remainder of this article is organized as follows. In Sec.~\ref{sec02}, we study the controlled-phase gate with an arbitrary phase, and then extend it to a CNOT gate. In Sec.~\ref{sec03}, we study the achievable gate fidelity with realistic parameters and fluctuation of qubit positions. Section~\ref{sec04} gives a comparison between the gate protocols here and previous Rydberg gates with distant qubits. A brief summary is given in Sec.~\ref{sec05}. 

\begin{figure}
\includegraphics[width=3.3in]
{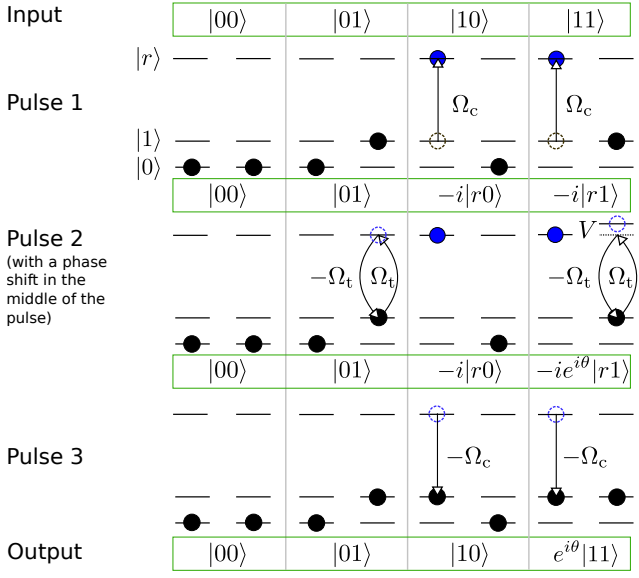}
\caption{A controlled-phase gate with an arbitrary phase $\theta$ mediated by a weak Rydberg interaction $V$ that satisfies the condition $\theta=-2\pi V/\sqrt{\hbar^2 \Omega_{\text{t}}^2+V^2}$~(up to a trivial addend of an integer times $2\pi$). The first and third pulses excite and deexcite the Rydberg excitation in the control qubit. At the middle of the second pulse a $\pi$ phase change is inserted in the Rabi frequency via modifying the phases in the Rydberg lasers. The phase $\theta$ is tunable via adjustment of the ratio between $\Omega_{\text{t}}$ and $V$.    \label{gatesequence} }
\end{figure}

\section{Two-qubit controlled gates with weak van der Waals interactions }\label{sec02}

\subsection{A controlled-phase gate with an arbitrary phase $\theta$ }\label{sec02A}
We first present the protocol for a controlled-phase gate and then extend it to a CNOT gate in Sec.~\ref{sec02B}. In particular, we consider a quantum entanglement operation 
\begin{eqnarray}
\{|00\rangle,|01\rangle, |10\rangle,|11\rangle\}  &\longmapsto & \{|00\rangle,|01\rangle,|10\rangle, e^{i\theta}|11\rangle\}, 
\label{gatemap}
\end{eqnarray}
by exciting two qubits to Rydberg states via external laser fields, where $\theta$ is tunable via adjustment of the frequency and magnitude of the laser field, and $|0\rangle$ and $|1\rangle$ are the two states of a qubit defined by the two hyperfine-Zeeman ground substates of a heavy alkali-metal atom such as $^{87}$Rb and $^{133}$Cs. External laser fields are sent to the qubits for exciting $|1\rangle$ to a Rydberg state $|r\rangle$. Equation~(\ref{gatemap}) is realized effectively by three pulses, where ``effectively'' means that the total number of pulses is four, but since the middle two pulses only differ by a phase which can be rapidly inserted in the laser fields, the total number of pulses is effectively three as shown in Fig.~\ref{gatesequence}. Pulses 1 and 3 are for exciting and deexciting the control qubits with Rabi frequencies of magnitude $|\Omega_{\text{c}}|$, and pulse 2 is for exciting the target qubit with Rabi frequencies of magnitude $|\Omega_{\text{t}}|$. Upon the application of pulse 1 with a $\pi$ rotation $|1\rangle\rightarrow|r\rangle$ for the control qubit enabled by the Hamiltonian $\Omega_{\text{c}}|r\rangle\langle1|/2+$H.c. in a rotating frame, the input eigenstates evolve as
\begin{eqnarray}
|01\rangle  &\longmapsto & |01\rangle,\nonumber\\
|10\rangle  &\longmapsto & -i|r0\rangle,\nonumber\\
|11\rangle  &\longmapsto & -i|r1\rangle, \nonumber
\end{eqnarray}
where the other input state $|00\rangle$ does not evolve since it is not excited because the energy separation between $|0\rangle$ and $|1\rangle$ is $E_{\text{hyper}}=h\times6.8$~GHz for $^{87}$Rb which is orders of magnitude larger than ($\hbar$ times) the Rydberg Rabi frequency $2\pi\times0.8$~MHz that we will consider. Then, the probability $2(\hbar\Omega_{\text{t}}/E_{\text{hyper}})^2$~\cite{Saffman2005} of Rydberg excitation of $|0\rangle$ is on the order of $10^{-8}$, which can be ignored. 

The Hamiltonian for pulse 2 sent upon the target qubit is $\pm\Omega_{\text{t}}|r\rangle\langle1|/2+$H.c., where the sign $+(-)$ applies for the first~(second) half of pulse 2. During the first half of pulse 2, laser fields are sent to the target qubit for the transition $|1\rangle\rightarrow|r\rangle$ with a Rabi frequency $\Omega_{\text{t}}$ for a pulse duration $t_{0}=2\pi/\sqrt{\Omega_{\text{t}}^2+V^2/\hbar^2}$, where $V$ is the Rydberg interaction of the state $|rr\rangle$. As previous studied in Refs.~\cite{Shi2017,Shi2018Accuv1,Shi2018prapp2,Levine2019}, the interaction $V$ will detune the transition between $|r1\rangle$ and $|rr\rangle$ so that the state rotation has a generalized Rabi frequency $\overline{\Omega_{\text{t}}}\equiv \sqrt{\Omega_{\text{t}}^2+V^2/\hbar^2}$. In contrast to the resonant Rabi rotation between two states where one full Rabi cycle results in a phase change $\pi$ to the state, the detuned Rabi cycle will lead to a phase change $\theta/2 = -\pi[1+ V/(\hbar\overline{\Omega_{\text{t}}})]$~\cite{Shi2017}, so that the state evolution in the first half of pulse 2 is
\begin{eqnarray}
|01\rangle  &\longmapsto &e^{-it_0 \hat{H}(\Omega_{\text{t}}) /\hbar} |01\rangle ,\nonumber\\
-i|r0\rangle  &\longmapsto & -i|r0\rangle,\nonumber\\
-i|r1\rangle  &\longmapsto & -ie^{i\theta/2}|r1\rangle,\nonumber
\end{eqnarray}
where
\begin{eqnarray}
 \hat{H}(\Omega_{\text{t}}) &=&\hbar[\Omega_{\text{t}} |0r\rangle\langle01|] /2+\text{H.c.} \nonumber
\end{eqnarray} 
The second half of pulse 2 is similar to its first half, but with a Rabi frequency $-\Omega_{\text{t}}$. With a radiation duration $t_0$, the input state $|01\rangle$ returns to itself since its total time-evolution operator in the second pulse $ e^{-it_0 \hat{H}(-\Omega_{\text{t}}) /\hbar}e^{-it_0 \hat{H}(\Omega_{\text{t}}) /\hbar}$ reduces to an identity because $\hat{H}(-\Omega_{\text{t}})+\hat{H}(\Omega_{\text{t}}) = 0$. On the other hand, remarkably, the change of the sign of the Rydberg Rabi frequency does not change the picture of the detuned Rabi oscillation, so that another $\theta/2$ phase change appears for the input state $|11\rangle$. As a consequence, the state evolution is
\begin{eqnarray}
e^{-it_0 \hat{H}(\Omega_{\text{t}}) /\hbar}|01\rangle  &\longmapsto & |01\rangle ,\nonumber\\
-i|r0\rangle  &\longmapsto & -i|r0\rangle,\nonumber\\
-ie^{i\theta/2}|r1\rangle  &\longmapsto & -ie^{i\theta}|r1\rangle, \label{eq05}
\end{eqnarray}
in the second half of pulse 2.

Pulse 3 is similar to pulse 1 but with a $\pi$ phase change to the Rabi frequency which becomes $-\Omega_{\text{t}}$. The $\pi$ rotation restores the state of the control qubit back to the ground state, leading to 
\begin{eqnarray}
|01\rangle  &\longmapsto & |01\rangle ,\nonumber\\
-i|r0\rangle  &\longmapsto & |10\rangle,\nonumber\\
-ie^{i\theta}|r1\rangle  &\longmapsto & e^{i\theta}|11\rangle.\nonumber
\end{eqnarray}
With the understanding that the input state $|00\rangle$ does not evolve in the rotating frame, the state transform in Eq.~(\ref{gatemap}) is realized as shown in Fig.~\ref{gatesequence}.
\begin{figure}
\includegraphics[width=3.3in]
{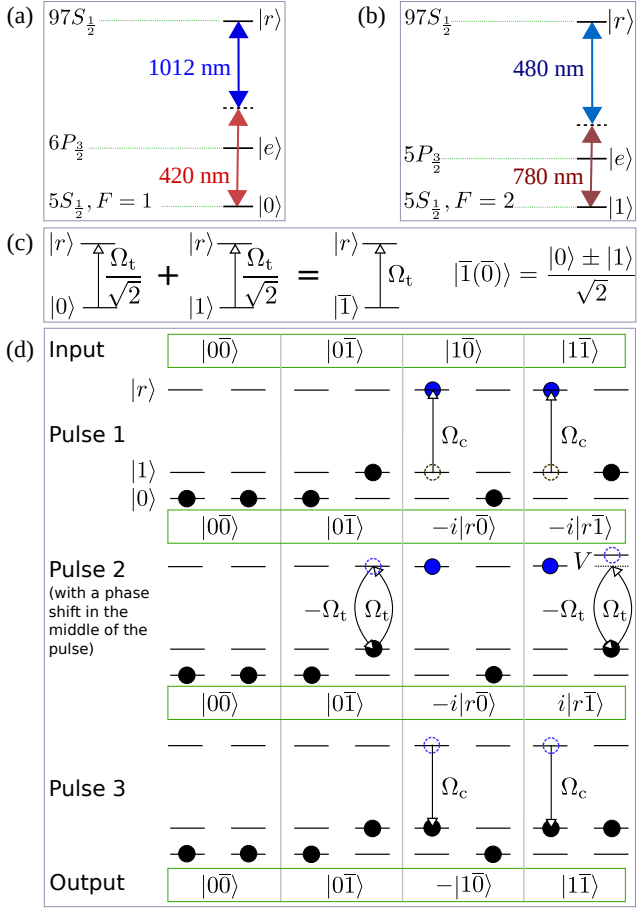}
\caption{A CNOT gate with a pulse sequence similar to that in Fig.~\ref{gatesequence} when $\theta=\pi$. (a,b) During pulse 2, both qubit states $|0\rangle$ and $|1\rangle$ are excited to the Rydberg state $|r\rangle$ via two different low-lying $p$-orbital states. (c) When the Rydberg Rabi frequencies for $|0\rangle\rightarrow|r\rangle$ and $|1\rangle\rightarrow|r\rangle$ are both $\Omega_{\text{t}}/\sqrt{2}$, it is like that the state $|\overline{1}\rangle\equiv(|0\rangle+|1\rangle)/\sqrt{2}$ is excited with a Rabi frequency $\Omega_{\text{t}}$ during pulse 2, while the other state $|\overline{0}\rangle\equiv(|0\rangle-|1\rangle)/\sqrt{2}$ stays intact. (d) By a similar pulse sequence as in Fig.~\ref{gatesequence} with the modification of pulse 2 shown in (c) with the condition $\hbar\Omega_{\text{t}}/V=\sqrt{3}$, a state mapping diag$\{1,1,-1,1\}$ is realized in the basis $\{|0\overline{0}\rangle, |0\overline{1}\rangle, |1\overline{0}\rangle, |1\overline{1}\rangle\}$, which is the standard CNOT gate after expanding $|\overline{0(1)}\rangle$. 
  \label{CNOTgate} }
\end{figure}

\subsection{A CNOT gate}\label{sec02B}
The controlled phase gate studied in Sec.~\ref{sec02A} becomes a C$_Z$ gate when $\theta=\pi$, which has the same entangling power~\cite{Williams2011} as a CNOT gate. On the other hand, the pulse sequence in Sec.~\ref{sec02A} can be slightly modified so as to directly create a CNOT gate,
\begin{eqnarray}
\{|00\rangle,|01\rangle, |10\rangle,|11\rangle\}  &\longmapsto & \{|00\rangle,|01\rangle,|11\rangle, |10\rangle\}. 
  \label{cnot}
\end{eqnarray}
The CNOT gate sequence shown in Fig.~\ref{CNOTgate} is similar to that described in Fig.~\ref{gatesequence} with two modifications. First, pulse 1 and pulse 3 use the same set of laser fields, i.e., no need to have a relative sign difference in the Rabi frequencies as in Fig.~\ref{gatesequence}. Second, in pulse 2, we replace the Rydberg-excitation Hamiltonian $\pm\hbar\Omega_{\text{t}}|r\rangle\langle1|/2+$H.c. in Fig.~\ref{gatesequence} by
\begin{eqnarray}
\pm\hbar\Omega_{\text{t}}|r\rangle(\langle0|+\langle1|)/(2\sqrt{2})+\text{H.c.},\label{pulse2H01}
\end{eqnarray}
 i.e., (i) the first half of pulse 2 requires two sets of laser fields, one for the Rydberg excitation $|0\rangle\xrightarrow[]{\Omega_{\text{t}}/\sqrt{2}}|r\rangle$, and the other for the Rydberg excitation $|1\rangle\xrightarrow[]{\Omega_{\text{t}}/\sqrt{2}}|r\rangle$, and (ii) the second half of pulse 2 requires two sets of laser fields, one for the Rydberg excitation $|0\rangle\xrightarrow[]{-\Omega_{\text{t}}/\sqrt{2}}|r\rangle$, and the other for the Rydberg excitation $|1\rangle\xrightarrow[]{-\Omega_{\text{t}}/\sqrt{2}}|r\rangle$. This requires simultaneous excitation of both qubit states to a Rydberg state. For two-photon Rydberg excitations, different intermediate states should be used with an example shown in Figs.~\ref{CNOTgate}(a) and~\ref{CNOTgate}(b). To understand the time evolution of the wavefunctions for different input states in pulse 2, we follow Ref.~\cite{Shi2020} and consider $\{|0\overline{0}\rangle, |0\overline{1}\rangle, |1\overline{0}\rangle, |1\overline{1}\rangle\}$, where $|\overline{1}(\overline{0})\rangle=(|0\rangle\pm|1\rangle)/\sqrt{2}$ as shown in Fig.~\ref{CNOTgate}(c). In this basis, Eq.~(\ref{pulse2H01}) becomes
\begin{eqnarray}
\hat{\bar{H}}(\pm\Omega_{\text{t}}) =\pm\hbar\Omega_{\text{t}}|r\rangle\langle\overline{1}|/2+\text{H.c.}\nonumber\label{pulse2H02}
\end{eqnarray}
With the condition $\hbar\Omega_{\text{t}}/V=\sqrt{3}$~(corresponding to $\theta=\pi$ in the picture of Sec.~\ref{sec02}), the first half and the second half of pulse 2, each with duration $t_{0}=\sqrt{3}\pi/\Omega_{\text{t}}$, lead to 
\begin{eqnarray}
|0\overline{1}\rangle  &\longmapsto &e^{-it_0 \hat{{\bar{H}}}(\Omega_{\text{t}}) /\hbar} |0\overline{1}\rangle \longmapsto  |0\overline{1}\rangle,\nonumber\\
-i|r\overline{0}\rangle  &\longmapsto & -i|r\overline{0}\rangle\longmapsto  -i|r\overline{0}\rangle,\nonumber\\
-i|r\overline{1}\rangle  &\longmapsto & |r\overline{1}\rangle\longmapsto i|r\overline{1}\rangle,\nonumber
\end{eqnarray}
for which a detailed proof is given above Eq.~(\ref{eq05}). Figure~\ref{CNOTgate}(c) shows that the Rydberg deexcitation via pulse 3 leads to a CNOT gate of Eq.~(\ref{cnot}) with the fact that 
\begin{eqnarray}
|\overline{1}\rangle \langle\overline{1}| -|\overline{0}\rangle \langle\overline{0}| & = & |0\rangle \langle1| + |1\rangle \langle 0|.\nonumber
\end{eqnarray}

\section{Gate fidelity}\label{sec03}
The gates can be fast. The total gate duration is $t_{\text{g}}=2\pi(1/\Omega_{\text{c}}+2/\overline{\Omega_{\text{t}}})$ according to the pulse sequence in Fig.~\ref{gatesequence} or Fig.~\ref{CNOTgate}. For a C$_Z$ or CNOT gate where $\theta=\pi$, we should have the condition 
\begin{eqnarray}
  \Omega_{\text{t}} = \sqrt{3}V/\hbar \label{Rabicondition01}
\end{eqnarray}
  so that the total gate duration is $t_{\text{g}}=2\pi(1/\Omega_{\text{c}}+\sqrt{3}/\Omega_{\text{t}})$. Rydberg Rabi frequencies around $2\pi\times4.6$~MHz were used for entanglement of individual neutral atoms in Ref.~\cite{Graham2019}, so the C$_Z$ or CNOT gates can have a short duration $t_{\text{g}}=0.59~\mu$s if similar Rabi frequencies are employed.

  A high fidelity~(about $99\%$) can be realized with our gates. Because the gate sequences for the C$_Z$ and the CNOT gates are similar, we consider the C$_Z$ gate as an example. In order to avoid slow gate speed, a Rydberg interaction around or over $h\times0.5$~MHz is desirable. Then, higher Rydberg states are preferred so that the Rydberg interaction can still be around $h\times0.5$~MHz even if the qubits are separated by tens of $\mu$m. Among the previous demonstrations of neutral atom quantum gates, the highest $d$-orbital Rydberg state of $^{87}$Rb atoms ever used had a principal quantum number $n=97$~\cite{Isenhower2010,Zhang2010}. The excitation of $s$-orbital Rydberg states with $n=102$ was demonstrated in an ensemble of $^{87}$Rb atoms where coherent many-body phenomena were observed~\cite{Dudin2012}. To avoid the anisotropic interaction of $d$-orbital Rydberg states, we consider $s$-orbital Rydberg states whose interaction is highly isotropic in real space~\cite{Walker2008}. For these reasons, we analyze the gate performance with $|r\rangle\equiv |97S_{1/2},m_J=1/2\rangle$ and Rydberg Rabi frequencies 
  \begin{eqnarray}
\Omega_{\text{c}}=\Omega_{\text{t}}=2\pi\times0.8~\text{MHz},\label{Rabifreq}
  \end{eqnarray}
  so that the duration will be $3.4~\mu$s for a C$_Z$ gate. We note that a similar Rydberg Rabi frequency $2\pi\times0.81$~MHz was used in~\cite{Zhang2010} for entanglement of individual $^{87}$Rb atoms via the excitation of the $97d$ states. The Rydberg interaction of $|rr\rangle$ for well separated qubits is characterized by $C_6/\mathcal{L}^6$~\cite{Walker2008,Saffman2010}, where $C_6 = h\times 39.5$~THz$\mu$m$^6$~\cite{Shi2014} is the van der Waals coefficient and $\mathcal{L}$ is the qubit spacing. To analyze the performance of a C$_Z$ gate where $\theta=\pi$ in Eq.~(\ref{gatemap}), the condition in Eq.~(\ref{Rabicondition01}) requires that the distance between the centers of the traps for the two qubits should be $20.99~\mu$m with Eq.~(\ref{Rabifreq})~[which leads to $V\approx h\times0.46$~MHz according to Eq.~(\ref{Rabicondition01})].

There are intrinsic errors from the Rydberg-state decay and the position fluctuations of the atomic qubits when the traps are turned off during the gate sequence~(there is also an error due to Doppler dephasing but it is not an intrinsic issue since it is removable by the methods in, e.g., Refs~\cite{Ryabtsev2011,Shi2020prapplied}). The former can be estimated as $E_{\text{decay}}=T_{\text{Ryd}}/\tau$, where $\tau$ is the lifetime of the Rydberg state and $T_{\text{Ryd}}$ is the total duration for the input state to stay in the Rydberg state averaged over the four input eigenstates,
\begin{eqnarray}
 T_{\text{Ryd}} &=& \frac{1}{4}\sum_{\alpha\in\{0,1\}} \sum_{|\psi(0)\rangle\in\{|01\rangle, |10\rangle, |11\rangle\}}\int \Big[ |\langle \alpha r|\psi(t)\rangle|^2
    \nonumber\\
    && + |\langle r\alpha |\psi(t)\rangle|^2+ 2|\langle rr|\psi(t)\rangle|^2 \Big]dt ,\nonumber%
\end{eqnarray}
where $|\psi(t)\rangle$ is the wavefunction calculated by unitary time evolution from the initial state $|\psi(0)\rangle$. Numerical simulation shows $T_{\text{Ryd}}\approx1.52\times2\pi/\Omega_{\text{c}}$ in general and $T_{\text{Ryd}}\approx1.91~\mu$s with Eq.~(\ref{Rabifreq}), and $\tau$ is $0.311$~ms~(or 1.10~ms) at room temperature~(or at $4.2$~K)~\cite{Beterov2009}, leading to $E_{\text{decay}}=6.14\times10^{-3}$~(or $1.74\times10^{-3})$. 

\begin{figure}
\includegraphics[width=3.3in]
{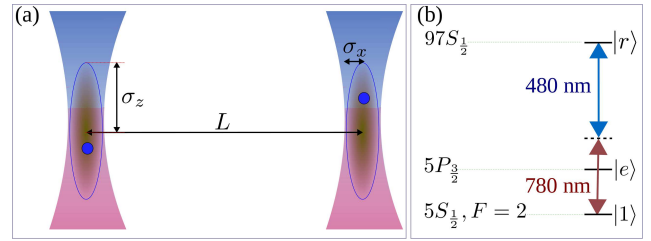}
\caption{(a) Schematic of the position fluctuation of the two atomic qubits in real space. The trap centers are at $(x_{\text{c}},y_{\text{c}},z_{\text{c}})$ and $(L+x_{\text{t}},y_{\text{t}},z_{\text{t}})$, while the locations of the qubits can deviate from the centers of the traps. The longitudinal and transverse r.m.s. position fluctuations of the atoms in the traps are $\sigma_z$ and $\sigma_\perp$ along the $\mathbf{z}$ and $\mathbf{x}$~(or $\mathbf{y}$) directions. The gate sequence proceeds after the traps are switched off so that the atoms fly freely. (b) Rydberg excitation of the qubit state $|1\rangle\equiv^{87}$Rb$|5S_{1/2},F=2,m_F=2\rangle$ via a largely detuned intermediate state $|e\rangle$. The red and blue shadows in (a) indicate the lower and upper laser fields in (b) used for exciting the Rydberg states from $|1\rangle$.    \label{position} }
\end{figure}

The fluctuation of the qubit positions contributes the major error because our gate works perfectly only when the interaction is equal to the desired value. There will be random position fluctuation of the qubits before the traps are switched off, and the atoms fly freely during the gate sequence. This leads to fluctuation of the qubit spacing which results in fluctuation of $V$. A numerical investigation of this matter requires information for realistic r.m.s. qubit-position fluctuations in reference to the trap center. Though most publications on entanglement experiments did not show such details, Ref.~\cite{Graham2019} showed that the longitudinal and transverse r.m.s. position fluctuations of the atoms in their traps were $\sigma_{z0}=1.47~\mu$m and $\sigma_{\perp0}=0.27~\mu$m, respectively. In order to incorporate the effect of the free flight of the atoms during the gate sequence, we would like to modify the values of r.m.s. fluctuations. The change of the atomic locations during the gate sequence will change the interatomic distance. For an atomic temperature around $T_a=10~\mu$K~(e.g., qubits with $T_a=5,~10,~15~\mu$K were studied in the entanglement experiments of Refs.~\cite{Picken2018},~\cite{Levine2018}, and~\cite{Graham2019}, respectively), the r.m.s. distance for a qubit to fly during the gate sequence is about $\ell= v_{\text{rms}}t_{\text{g}}$, where $v_{\text{rms}}= \sqrt{k_BT_a/m}$ is the r.m.s. speed of the atom along one of the three directions in the 3D space, $T_a$ is the effective temperature of the atom, and $k_B$ and $m$ are the Boltzmann constant and the atomic mass, respectively. For a free flight of qubits, the average change of distance will be $v_{\text{rms}}t_{\text{g}}/2$ for a duration $t_{\text{g}}$. So, we analyze the fluctuation of the qubit positions by increasing the r.m.s. fluctuations to 
\begin{eqnarray}
 \sigma_{z}&=&\sigma_{z0}+\ell/2,\nonumber\\
 \sigma_{\perp}&=&\sigma_{\perp0}+\ell/2,\nonumber%
\end{eqnarray}
which leads to $(\sigma_{z}, \sigma_{\perp})=(1.52,~0.32)~\mu$m in the condition of Eq.~(\ref{Rabifreq}). 

The schematic of position fluctuation is shown in Fig.~\ref{position}(a). We suppose that the quantization axis is along $\mathbf{x}$, and $\pi$ polarized Rydberg laser lights travel along $\mathbf{z}$. The Rydberg excitation scheme is shown in Fig.~\ref{position}(b). The centers of the traps for the control and target qubits are $(0,0,0)$ and $(L,0,0)$, respectively. Because of the finiteness of the trap depths, the positions of the control and target qubits are $(x_{\text{c}},y_{\text{c}},z_{\text{c}})$ and $(L+x_{\text{t}},y_{\text{t}},z_{\text{t}})$, respectively, where the distribution function
\begin{eqnarray}
\mathscr{G}(\zeta) &=& \frac{1}{\sqrt{2\pi} \sigma_\perp}e^{-\zeta^2/(2\sigma_\perp^2)}\nonumber
\end{eqnarray}
characterizes $\zeta\in\{x_{\text{c}},y_{\text{c}},x_{\text{t}},y_{\text{t}}\}$, and 
\begin{eqnarray}
\mathscr{G}(\zeta) &=& \frac{1}{\sqrt{2\pi} \sigma_z}e^{-\zeta^2/(2\sigma_z^2)}\nonumber
\end{eqnarray}
is the distribution function for $\zeta=z_{\text{c}}$ or $z_{\text{t}}$. In the numerical simulation, the value of $V$ is calculated by using the actual distance of the qubits $\mathcal{L}=$$\sqrt{(x_{\text{c}}-x_{\text{t}}-L)^2 +(y_{\text{c}}-t_{\text{t}})^2 +(z_{\text{c}}-z_{\text{t}})^2 }$. The average fidelity is sampled with 
\begin{eqnarray}
\overline{\mathcal{F}}&=&\Pi_{\zeta}\int\mathscr{G}(\zeta)d\zeta\mathcal{F},\nonumber%
\end{eqnarray}
where the fidelity $\mathcal{F}$ is defined by~\cite{Pedersen2007}
\begin{eqnarray}
\mathcal{F} &=&\left[  |\text{Tr}(U^\dag \mathscr{U})|^2 + \text{Tr}(U^\dag \mathscr{U}\mathscr{U}^\dag U ) \right]/20.\nonumber
\end{eqnarray}
Here, $\mathscr{U}$ is the actual gate matrix evaluated by using the unitary dynamics with the Rydberg-state decay ignored, and $U$=diag$\{1,1,1,-1\}$ is the ideal gate matrix in the basis $\{|00\rangle,|01\rangle,|10\rangle,|11\rangle\}$. The integral for each $\zeta$ is approximated by taking discrete values; for example, for $x_{\text{c}}$, the sampling is approximated by $\sum \mathscr{G}(x_{\text{c}}) \mathcal{F}/\sum \mathscr{G}(x_{\text{c}})$ with $x_{\text{c}}\in\{-1.5,-1.5+\delta,-1.5+2\delta,~\cdots,~1.5-2\delta,1.5-\delta,1.5\}\sigma_\perp$, and the total gate fidelity is sampled by similar approximations for $\{y_{\text{c}},z_{\text{c}},x_{\text{t}},y_{\text{t}},z_{\text{t}}\}$. The numerical simulation is resource-demanding, but we can obtain an estimate about the convergence of the integration via slowly varying the steps. For $\delta$ equal to $0.25,0.2,0.15,0.12$, and $0.1$, the sampled average fidelities are $0.9910,0.9912,0.9914,0.9920$, and $0.9920$ respectively, from which we estimate $\overline{\mathcal{F}}=0.992$. With Rydberg-state decay included, the intrinsic fidelity would be  $\overline{\mathcal{F}}-E_{\text{decay}}=0.986$~(or $0.990$) at $300$~K~(or 4.2~K).

\section{Comparison with other methods}\label{sec04}
The gates in this work have several advantages compared to other gates~\cite{PhysRevLett.85.2208,Shi2017,Shi2018Accuv1,Beterov2016} that can also work with weak interactions in distant Rydberg atoms.

The first advantage is that our gate does not have rotation errors in the weak interaction regime. It was proposed in Ref.~\cite{PhysRevLett.85.2208} that phase gates can be created via the dynamical phase shift of the van der Waals interaction between two weakly interacting Rydberg atoms as experimentally demonstrated in Ref.~\cite{Jo2019}. The method of~\cite{PhysRevLett.85.2208} involves the excitation $|11\rangle\rightarrow|rr\rangle$ which has an intrinsic error $2V^2/(\hbar^2\Omega^2)$ due to the failure for exciting the state $|rr\rangle$ from $|11\rangle$ as analyzed in Ref.~\cite{Saffman2005}. In contrast, the intrinsic accuracy of the gate here is limited only by the Rydberg-state decay in the weak interaction regime. 

The second advantage is that our gate depends on van der Waals interaction that naturally appears between two distant Rydberg atoms. As has been experimentally tested~\cite{Jo2019}, such a strategy allows an easy access to an experimental demonstration. In comparison, by using F\"{o}rster resonance one can also design entangling gates between well-separated qubits~\cite{Beterov2016,Beterov2020,Beterov2018,Beterov2018arX} but fine tuning via external fields is required~\cite{PhysRevLett.47.405,PhysRevLett.80.249,PhysRevLett.80.253,Westermann2006,PhysRevLett.108.113001,PhysRevA.93.042505,PhysRevA.93.012703,Ryabtsev2010,Tretyakov2014,Faoro2015,Tretyakov2017,Cheinet2020}. However, it will be useful to investigate, if possible, an extension of our method to the regime of F\"{o}rster resonance. If it is indeed possible, high-fidelity entanglement should be reachable with qubits separated even longer than the estimate in this work. This is because the F\"{o}rster resonant interaction falls off as $1/\mathcal{L}^3$, while the van der Waals interaction falls off as $1/\mathcal{L}^6$ when $\mathcal{L}$ increases. 

The third advantage is that our method can more easily realize a C$_Z$ gate compared to previous high-fidelity gates based on weak van der Waals interactions~\cite{Shi2017,Shi2018Accuv1}. Compared to Ref.~\cite{Shi2017} which uses five pulses for coupling two different Rydberg states, the gate here needs only three steps to couple one Rydberg state. Compared to~\cite{Shi2018Accuv1} which proposes a quantum entangling gate that should be repeated several times to form a C$_Z$ gate when assisted by single-qubit gates, our method can realize a controlled-phase gate~(such as C$_Z$) with any desired phase by only three laser pulses as shown in Fig.~\ref{gatesequence}.

\section{conclusions}\label{sec05}
We study a controlled-phase gate based on weak van der Waals interactions between neutral Rydberg atoms. The gate is realized via phase accumulations in detuned Rabi oscillations enabled by weak Rydberg interactions. The gate is easily realizable and can have a high fidelity because it only needs three steps to couple one type of Rydberg state, has no intrinsic rotation errors in the weak interaction regime, and does not depend on fine tuning of F\"{o}rster resonant processes. We use practical parameters to show that it is possible to create a C$_Z$ gate in two qubits separated by $21~\mu$m with a fidelity about $0.990~($or $0.986)$ at 4.2~K~(or at room temperature). We also show that a similar pulse sequence can lead to a CNOT gate which is realizable with similarly small van der Waals interactions. The theory brings hope to simplify the quantum circuit for large-scale quantum computing in which entanglement between distant qubits is required.

\section*{ACKNOWLEDGMENTS}
This work is supported by the National Natural Science Foundation of China under Grants No. 12074300 and No. 11805146, the Natural Science Basic Research plan in Shaanxi Province of China under Grant No. 2020JM-189, and the Fundamental Research Funds for the Central Universities.

%


\end{document}